# Bacterial nucleoid: Interplay of DNA demixing and supercoiling

Marc Joyeux[(#)]

*Laboratoire Interdisciplinaire de Physique,*

*CNRS and Université Grenoble Alpes,*

*Grenoble, France*

**Running title**: DNA demixing and supercoiling.

**Abstract:** This work addresses the question of the interplay of DNA demixing and supercoiling in bacterial cells. Demixing of DNA from other globular macromolecules results from the overall repulsion between all components of the system and leads to the formation of the nucleoid, which is the region of the cell that contains the genomic DNA in a rather compact form. Supercoiling describes the coiling of the axis of the DNA double helix to accommodate the torsional stress injected in the molecule by topoisomerases. Supercoiling is able to induce some compaction of the bacterial DNA, although to a lesser extent than demixing. In this paper, we investigate the interplay of these two mechanisms, with the goal of determining whether the total compaction ratio of the DNA is the mere sum or some more complex function of the compaction ratios due to each mechanism. To this end, we developed a coarse-grained bead-and-spring model and investigated its properties through Brownian dynamics simulations. This work reveals that there actually exist different regimes, depending on the crowder volume ratio and the DNA superhelical density. In particular, a regime where the effects of DNA demixing and supercoiling on the compaction of the DNA coil simply add up is shown to exist up to moderate values of the superhelical density. In contrast, the mean radius of the DNA coil no longer decreases above this threshold and may even increase again for sufficiently large crowder concentrations. Finally, the model predicts that the DNA coil may depart from the spherical geometry very close to the jamming threshold, as a trade-off between the need to minimize both the bending energy of the stiff plectonemes and the volume of the DNA coil to accommodate demixing.

[(#)] marc.joyeux@univ-grenoble-alpes.fr

**Statement of significance:** Many biological processes take place simultaneously in living cells. It is tempting to study each of them separately and rely on the hypothesis that cells behave like the "addition" of the isolated parts. This is however not always the case and the present work illustrates this fact. We consider two different processes, which are both able to compact the bacterial DNA, namely demixing and supercoiling, and we study how the DNA reacts when subject to both of them simultaneously. Through coarse-grained modeling and Brownian dynamics simulations, we show that the two processes are "additive" only in a limited range of biologically relevant values of the parameters and that their interplay is much more complex outside from this range.

# INTRODUCTION

This work addresses the question of the interplay of DNA demixing and supercoiling in bacterial cells. Prokaryotic cells are simpler than eukaryotic ones in many respects, yet several of their fundamental features remain poorly understood. For example, the mechanism leading to the formation of the bacterial nucleoid is a longstanding but still lively debated question (1-5). The nucleoid is the region of the cell which contains the genomic DNA, together with a certain number of proteins and other macromolecules (6). In contrast with the nucleus of eukaryotic cells, the bacterial nucleoid is not separated from the rest of the cytosol by a bounding membrane. It nevertheless occupies only a fraction of the cell, whose volume depends sensitively on several factors, like the richness of the nutrients (7-11), the cell cycle step (12,13), and the eventual addition of antibiotics (9-11,14-18). This is quite puzzling, because the volume of the unconstrained bacterial genomic DNA in physiological saline conditions (estimated from the Worm-Like Chain (WLC) model (19)) is approximately thousand times larger than the volume of the cell. One has therefore to understand why the DNA molecule remains localized inside the nucleoid instead of expanding throughout the cell. It is becoming increasingly clear that the formation of plectonemes, the bridging of DNA duplexes by nucleoid proteins, and the action of short-range attractive forces, which are commonly evoked as the mechanisms responsible for the formation of the bacterial nucleoid (1), may not play the leading role in the compaction of the DNA (see for example (4) and references therein). In contrast, the 20 years old proposition that increasing amounts of non-binding globular macromolecules may be able to compact the genomic DNA gradually (20-24) has recently received strong support, both from the experimental (25-27) and computational (28-33) sides. The proposed mechanism is that the overall repulsion between all components of the system leads to a separation into two phases (34-45), one of them being rich in DNA and poor in the other macromolecule (the nucleoid) and the other one being almost deprived of DNA (the rest of the cytosol). The connectivity of the long DNA molecule and its ability to deform close to and around the proteins to fit in void spaces between proteins play crucial roles in this mechanism, in that they induce many-body interaction terms that ultimately result in effective DNA-DNA and protein-protein attraction when DNA is depleted from the regions between particles. Compaction of the genomic DNA through its demixing from other macromolecules of the cytosol is the first point this work focuses on.

The second point deals with supercoiling, that is the winding about itself of the circular DNA double-helix in response to the torsional stress induced by topoisomerases (46).

Experimental and theoretical results suggest that supercoiling is able to induce some limited compaction of the bacterial DNA. For example, relaxation of the underwinding of the circular DNA molecule through inhibition of the topoisomerase activity leads to a modest increase in the size of *E. coli* nucleoids (14,47). Moreover, theoretical arguments suggest that the radius of gyration of an unconstrained supercoiled DNA molecule with contour length of 2.6 mm is of the order of 2.5 µm (48), which is smaller than the WLC estimate for a circular chain with the same contour length (about 3.4 µm), but still significantly larger than the average dimensions of *E. coli* cells.

The question addressed in this paper is that of the interplay of these two mechanisms, DNA demixing and supercoiling, which are both able to compact the bacterial DNA, and more precisely the question whether the total compaction of the DNA coil is the mere sum or some more complex function of the compaction ratios due to each mechanism. Stated in other words, is the increase of DNA compaction ratio provoked by an increase of crowder density similar for torsionally relaxed (less compact) and supercoiled (more compact) DNA coils ? Or, conversely, is the increase of DNA compaction ratio provoked by an increase of torsional stress similar for DNA immersed in a dilute cytosol (less compact DNA coils) and a highly crowded cytosol (more compact DNA coils) ? Indeed, plectonemes are composed of two intertwined DNA duplexes and are consequently thicker (estimated diameter in the range 10-32 nm for standard values of the underwinding of *in vivo* DNA (48-50) and more rigid (estimated persistence length of ≈80 nm (48)) than simple duplexes (diameter of ≈2 nm and persistence length of ≈50 nm). One may therefore expect that the demixing mechanism is less efficient in compacting plectonemic DNA than linear DNA, which implies that the effects of the two mechanisms do not simply add up.

More generally, understanding the interplay of DNA demixing and supercoiling is important for rationalizing some *in vivo* observations. For example, the nucleoid of certain bacteria, like *E. coli*, is divided into different macro-domains (4 macro-domains for *E. coli* cells), which display quite different densities of DNA nucleotides (51-54). It is known that certain families of nucleoid proteins probably contribute to the organisation of these macro-domains (55-57). On the other hand, it is also known that the DNA molecule is dynamically partitioned into several hundreds of independently supercoiled loops with average size ≈10 kb, which are called topological domains (58,59). One may therefore reason that the difference in DNA density in different macro-domains may result from different levels of supercoiling in the corresponding topological domains, provided that different levels of supercoiling result in different levels of DNA compaction at constant crowder density.

The topology and dynamics of free supercoiled DNA have received much attention from the experimental (47,49,60-64), theoretical (65), and numerical (66-72) points of view. The effects of confinement (73-75) and increasing nucleic acid concentration (72,76) on supercoiled DNA have also been investigated. In contrast, much less work has been devoted to the influence of non-binding globular macromolecules on the conformations of supercoiled DNA (22,48,77,78). In the perspective of the present work, the most striking result is probably the observation that in crowded conditions the size of supercoiled DNA may exceed that of its linear variant (77), which supports the putative non-additivity of the two compaction mechanisms. Moreover, condensation experiments suggest that the tight packing of DNA supercoils in condensates is facilitated by the decrease of the diameter of plectonemes rather than by a variation of the writhe/twist ratio (78), which points towards the need for mechanical rearrangements of the plectonemes to accommodate strong compaction ratios. These results were however obtained for short plasmids (less than 3000 base pairs) in free solution and need to be confirmed for longer and confined DNA molecules.

In order to shed light on the interplay of DNA demixing and supercoiling, a coarse-grained model was developed along the same lines as those used previously to investigate facilitated diffusion (79-81), the interactions of DNA and H-NS nucleoid proteins (82-84), and the formation of the bacterial nucleoid (4,5,32,33,85). Torsional energy was accounted for in the model as described in (86) and the properties of the full model were investigated for different values of the number of crowders and the superhelical density (*i.e.* the relative overwinding) of the DNA chain. In particular, the number of crowders was increased up to the jamming threshold, where strong compaction is known to occur (32,33), and the investigated range of superhelical density values encompasses the estimated value for *E. coli* (87). The results presented in this article reveal that there actually exist different regimes, which are separated by threshold values of the crowder volume ratio and the DNA superhelical density. In particular, a regime where the effects of DNA demixing and supercoiling on the compaction of the DNA coil simply add up is shown to exist up to moderate values of the superhelical density, while the mean radius of the DNA coil ceases to decrease above this threshold and may even increase again for sufficiently large crowder concentrations. Moreover, the model predicts that the DNA coil may depart from the spherical geometry very close to the jamming threshold, as a trade-off between the need to minimize the bending energy of stiff plectonemes and the need to minimize the volume of the DNA coil to accommodate demixing.

**METHODS**

The coarse-grained bead-and-spring model developed for the present study is described in detail in Model and Simulations in the Supporting Material. In brief, the DNA molecule is modeled, as in (84), as a circular chain of 2880 beads with radius 1.0 nm separated at equilibrium by a distance 2.5 nm and enclosed in a confinement sphere of radius 120 nm, which represents the cell envelope. Two beads represent 15 base pairs and the circular chain therefore corresponds to 21600 base pairs. Both the contour length of the DNA chain and the volume of the confinement sphere are about 200 times smaller than their actual values in *E. coli* cells, so that the nucleic acid concentration of the model is close to the *in vivo* concentration of most bacteria ($\approx$10 mM). Note that the spherical confinement chamber is adequate for modeling cocci, but not bacilli (including *E. coli*), which look rather like capped cylinders. DNA beads interact through stretching, bending, torsional, and electrostatic terms. The bending rigidity constant is chosen so that the model reproduces the known persistence length of double-stranded DNA ($\approx$50 nm). The torsional energy term is borrowed from (86) and the torsional rigidity is adjusted so that at equilibrium the writhe contribution accounts for approximately 70% of the linking number difference (49). The values of the bending and torsional rigidities are close together, in agreement with experimental results (49). Electrostatic repulsion between DNA beads is written as a sum of Debye-Hückel terms, which depend on effective electrostatic charges placed at the center of each bead. The values of these charges are derived from the net linear charge density along a DNA molecule immersed in a buffer with monovalent cations according to Manning's counterion condensation theory (88,89). The value of the Debye length ($\approx$1 nm) corresponds to a concentration of monovalent salt of 100 mM, which is the value that is generally assumed for the cytoplasm of bacterial cells.

Globular macromolecular crowders are modeled as a variable number $N$ of spheres with radius 7.4 nm and the same electrostatic charge as DNA beads. DNA-crowder and crowder-crowder interactions are expressed as sums of Debye-Hückel potentials, so that all components of the systems repel each other. It was shown previously (32,33) that maximum compaction of torsionally relaxed DNA chains is obtained when $\rho$, the effective crowder volume fraction (Eq. (S6)), is close to the jamming threshold for hard spheres, that is $\rho \approx 0.65$.

The properties of the present model for supercoiled DNA were investigated by integrating numerically overdamped Langevin equations with time steps of 10 ps for different values of the number of crowders, $N$, and the superhelical density of the DNA chain, $\sigma$. Various sets of trajectories were run with $N = 0$, 1500, 1750, 1875, and 2000, which corresponds to effective crowder volume ratios $\rho = 0$, 0.49, 0.57, 0.61, and 0.65, respectively. $\sigma$ was varied from 0 (torsionally relaxed DNA) to -0.08, thus encompassing the estimated value for *E. coli* cells, $\sigma \approx -0.06$ (87). Simulations were performed by first letting DNA chains with prescribed values of $\sigma$ equilibrate for 2 ms inside the confinement sphere in the absence of any crowder. The $N$ crowding spheres were then added at random, homogeneously distributed, and non-overlapping positions, and the complete system was allowed to equilibrate again for 5 ms ($N = 1500$ and 1750), 10 ms ($N = 1875$), or 20 ms ($N = 2000$), in order to cope with the marked slowing down of the dynamics close to the jamming threshold. The radius of the DNA coil, $R$ (computed as the mean distance of the DNA beads from the center of the confinement sphere), the excess of twist, $\Delta \text{Tw}$, and the writhe, $\text{Wr}$, were then averaged over time windows of 8 ms ($N = 0$), 15 ms ($N = 1500$ and 1750), 20 ms ($N = 1875$), and up to 60 ms for $N = 2000$. For the sake of better statistics, all results were moreover averaged over 4 different trajectories with different initial conditions but identical values of $N$ and $\sigma$. The error bars shown in the figures correspond to the standard deviation of the 4 average values obtained from the different trajectories with identical values of $N$ and $\sigma$. Temperature $T$ was assumed to be 298 K throughout the study. Representative snapshots extracted from trajectories with $N = 0$ and $N = 1875$ ($\rho = 0.61$) are shown in Figs. 1 and 2, respectively.

**RESULTS AND DISCUSSION**

The goal of this paper is to shed light on the interplay of DNA supercoiling and DNA demixing from other macromolecules with respect to the compaction of the DNA coil. To this end, several sets of trajectories were launched with different values of the effective crowder volume ratio, $\rho$, and the superhelical density of the DNA, $\sigma$. It is reminded that the superhelical density is defined according to $\sigma = \Delta Lk / Lk_0$, where $Lk_0$ and $Lk_0 + \Delta Lk$ are the linking numbers of the torsionally relaxed and topologically constrained DNA, respectively. The linking number difference $\Delta Lk$ is the sum of two contributions, namely the excess of twist around the DNA axis, $\Delta Tw$, and the writhe, $Wr$, which quantifies the winding of the DNA axis around itself. The linking number difference $\Delta Lk = \Delta Tw + Wr$ and the superhelical density $\sigma = \Delta Lk / Lk_0$ remain constant as long as the DNA molecule is not nicked. In contrast, the excess of twist $\Delta Tw$ and the writhe $Wr$ do fluctuate under the influence of thermal noise and external constraints. Note that in this work $\Delta Tw$, $Wr$, $\Delta Lk$, and $\sigma$, are all negative quantities, as a consequence of the underwinding of bacterial DNA, and that negative values of $\Delta Tw$ must be understood as a deficit of twist compared to torsionally relaxed DNA.

**The effect on DNA compaction of each mechanism taken separately**

The evolution of the compaction of the DNA coil with $\rho$ and $\sigma$ is summarized in Fig. 3, which shows the evolution of $<R>$ with $-\sigma$ for different values of $\rho$. The blue dashed line with open circular symbols indicates that the mean radius of the DNA coil decreases from 82 nm down to 75 nm when $|\sigma|$ is increased from 0 up to 0.08 in the absence of any crowder ($\rho = 0$). Along this curve, compaction results from the reorganization of the DNA chain, which forms an increasing number of plectonemes to reduce torsional stress (see Fig. 1). Plectonemes can conveniently be sought for as illustrated in Fig. S1, which displays the index $j$ of the bead located closest to bead $i$ for the equilibrated DNA conformations shown in Fig. 1. The nearest neighbors were searched for with the constraint that $j \notin [i-10, i+10]$, in order that the algorithm does not systematically select immediate neighbors along the DNA chain. In the vignette for torsionally relaxed DNA ($\sigma = 0$), the points are essentially randomly distributed, except for a discrete accumulation of points close to the diagonal, which correspond to the trivial case where the search algorithm led to $j = i \pm 11$. In the vignettes for $\sigma = -0.052$ and $\sigma = -0.078$, plectonemes appear instead as well defined segments parallel to the anti-diagonal. Plectonemes also appear in the vignette for $\sigma = -0.027$, but the segments are significantly shorter, thereby pointing towards a poor plectonemic structure. As illustrated in the top plot of Fig. 4 (blue dashed line with open circular symbols), and in agreement with previous work, the mean distance $<d>$ between the opposite strands of the plectonemes decreases significantly, like $1/|\sigma|$ (49), with increasing values of $|\sigma|$. Moreover, the mean contribution of the writhe to the linking number difference, $<Wr/\Delta Lk>$, decreases slightly with increasing values of $|\sigma|$ (blue dashed line with open circular symbols in the bottom plot of Fig. 4). It was shown previously that $<Wr/\Delta Lk>$ increases with $|\sigma|$ for short DNA sequences (300–3500 bp) and remains nearly constant for a 7 kbp DNA sequence (66). The present work therefore confirms that $<Wr/\Delta Lk>$ varies in opposite directions as a function of $|\sigma|$ for short (< 10 kbp) and long (> 10 kbp) DNA sequences.

On the other hand, the points on the ordinate axis ($\sigma = 0$) in Fig. 3 indicate that the mean radius of the torsionally relaxed DNA coil decreases from 82 nm down to 72.3 nm when $\rho$ is increased from 0 to 0.65. Along this axis, compaction of the DNA coil results from the demixing of the DNA chain and the spherical crowders, which are expelled outside from the DNA coil (32,33) (see Fig. 2). As was already observed in previous studies based on somewhat different models (32,33), demixing increases strongly close to the jamming threshold for spherical crowders ($\rho \approx 0.65$). This is more clearly seen in Fig. S2, which shows the evolution of $<R>$ with $\rho$ for torsionally relaxed DNA. A further slight increase of $\rho$ beyond 0.65 will probably result in significantly stronger compaction of the DNA coil, as in (32,33), but the dynamics of the system becomes too slow to be numerically tractable with our computer facility.

The purpose of the present paper is to decipher the rest of Fig. 3, that is to rationalize the compaction of topologically constrained DNA with increasing crowder volume fraction.

**The additive regime at moderate superhelical density**

A first remarkable feature of Fig. 3 is that the curves corresponding to different values of $\rho$ are parallel to the curve for $\rho = 0$ up to $-\sigma = 0.027$. This indicates that the effects of the two compaction mechanisms are actually simply additive for such moderate values of the superhelical density. As a consequence, the mean radius of the DNA coil is as small as 68.6 nm for $\rho = 0.61$ and $\sigma = -0.027$. The reason for such additivity is that the plectonemes are still few and loose and do not oppose compaction. This point can be checked in the vignettes for $\sigma = -0.027$ in Figs 1 and 2, where plectonemes are hardly noticeable, and Fig. S1, where the segments of points which signalize plectonemes are significantly shorter than for larger values of $|\sigma|$. Moreover, as was anticipated on the basis of theoretical grounds in Ref. (78) and observed upon increasing nucleic acid concentration in Ref. (77), the compaction of the DNA coil is facilitated by a decrease of the diameter of the superhelix with increasing crowder concentration. Indeed, the top plot of Fig. 4 indicates that for $\sigma = -0.027$ the mean distance $<d>$ between the opposite strands of the plectonemes decreases from 14.4 nm in the absence of any crowder down to 10.0 nm for $\rho = 0.61$. The fact that compaction of the DNA coil upon increase of the concentration of crowders is accompanied by an increase of $<Wr/\Delta Lk>$ (see bottom plot of Fig. 4) is more surprising, because one would rather imagine that compaction is favoured by unwinding rather than winding of the supercoils (78). A possible interpretation of this observation is that further winding of the supercoils from $<Wr/\Delta Lk> = 0.72$ ($<Wr> \approx -40$) to $<Wr/\Delta Lk> = 0.79$ ($<Wr> \approx -44$) contributes to the reduction of the diameter of the superhelix, which in turn facilitates compaction.

**The non-additive regime at larger superhelical density**

Additivity of the effects of DNA supercoiling and demixing on the compaction of DNA coils does however not extend to values of $|\sigma|$ larger than 0.027. This can be checked in Fig. 3, where the curves corresponding to different values of $\rho$ do not remain parallel to the curve for $\rho = 0$ above $|\sigma| = 0.027$. Owing to the uncertainties on the computed values of $<R>$ ($\approx 1$ nm), all what can be said safely is that (i) at moderate crowder volume ratios ($\rho = 0.49$ and $\rho = 0.57$), winding of the plectonemes beyond $|\sigma| = 0.027$ does not result in a significant increase of the compaction of the DNA coil, in contrast with the case $\rho = 0$, and (ii) closer to the jamming threshold ($\rho = 0.61$), winding of the plectonemes beyond $|\sigma| = 0.027$ even provokes significant decompaction of the DNA coil. This latter point is surprising, because it contrasts with the monotonous behavior of $<d>$ and $<Wr/\Delta Lk>$ with respect to $|\sigma|$ and $\rho$, which is observed in Fig. 4. Indeed, it was suggested above that a decrease in $<d>$ and an increase in $<Wr/\Delta Lk>$ facilitate compaction of the DNA coil. Then, why does the DNA coil expand for $\rho = 0.61$ and $|\sigma| > 0.027$, although $<d>$ goes on decreasing and $<Wr/\Delta Lk>$ goes on increasing ? As anticipated in the Introduction, the reason is that increasing $|\sigma|$ not only reduces the diameter of plectonemes, which makes compaction of the DNA coil easier, but also increases their stiffness, which has the inverse effect of opposing compaction. The increase in the stiffness of plectonemes as a function of $|\sigma|$ was checked for short DNA chains with 200 beads (1500 bp) in the absence of any crowder. These chains are sufficiently short for branching of plectonemes not to occur and a single plectoneme to be observed at any time. At regular time intervals, the two extremities of the plectoneme were sought for and the mid curve of the two opposite strands was computed. The directional correlation function of the segments of the mid curve was then averaged over many different conformations to get an estimate of the persistence length according to standard fitting procedures. The Log of the directional correlation function is shown for different values of $|\sigma|$ in the main plot of Fig. S3, and the evolution of the fitted values of the persistence length $\xi$ as a function of $|\sigma|$ in the insert of the same figure. It is seen that the persistence length (and consequently the formal bending rigidity) of the mid curve of the plectonemes increases by more than 50% from $|\sigma|=0.035$ to $|\sigma|=0.077$. All in all, Fig. 3 indicates that compaction of the DNA coil for $0.49 \leq \rho \leq 0.61$ and $|\sigma|>0.027$ results from the balance of two conflicting trends, namely the decrease of the diameter of the plectonemes and the increase of their stiffness, with the increase in stiffness becoming predominant at larger crowder concentrations.

**Departure from the spherical geometry at the jamming threshold**

Finally, for topologically constrained DNA, a third regime is observed at crowder concentrations very close to the jamming threshold. Indeed, in all the simulations with $\sigma=0$ or $\rho \leq 0.61$ the DNA chain relaxed towards a nearly spherical coil. This is illustrated in Fig. 5, which displays representative equilibrated conformations obtained with $\sigma=-0.078$ and $\rho=0.61$ (first row), and $\sigma=0$ and $\rho=0.65$ (second row). In the left (respectively, right) vignettes, the DNA coil is viewed parallel (respectively, perpendicular) to its principal axis of inertia with largest momentum. Because of the approximate spherical geometry, the aspect of the DNA coil does not change significantly from one viewpoint to the other one. Such nearly spherical DNA coils were also systematically observed in previous studies based on a somewhat different model of torsionally relaxed circular DNA (32,33). In contrast, for topologically constrained DNA ($|\sigma| \geq 0.027$) close to the jamming threshold ($\rho=0.65$), we observed in the present work that trajectories may relax either towards nearly spherical DNA coils or a fundamentally different type of nearly toroidal coils. This is illustrated in Fig. 5, where the third and fourth rows display representative equilibrated conformations obtained with $\sigma=-0.065$ and $\rho=0.65$. While in the third row the geometry of the DNA coil is again nearly spherical, so that the coil as the same aspect when viewed from any axis, this is no longer the case in the fourth row, where the geometry of the DNA coil is nearly toroidal, so that the coil appears like a ring when viewed parallel to the axis with largest momentum and like a disk when viewed perpendicular to this axis. As far as we can tell from 80 ms trajectories, both the spherical and toroidal conformations are stable or at least metastable. Since the toroidal geometry was never observed for torsionally relaxed DNA, it is most likely that the probability for the DNA chain to relax towards the toroidal geometry increases with $|\sigma|$. The rationale behind this observation is most probably that the toroidal geometry reduces the bending energy of the DNA coil by allowing the DNA chain to form large loops with radius close to that of the confinement sphere, while still allowing demixing and compaction perpendicular to the plane of the torus. It is therefore expected that increasing $|\sigma|$, thereby making plectonemes more rigid, should favor relaxation towards the toroidal geometry. Quite importantly, the mean radius of equilibrated DNA coils with toroidal geometry is of the order of 85-90 nm, which is larger than the mean radius of torsionally relaxed DNA coils in the absence of any crowder (about 82 nm, see Fig. 3). Still, such an increase in the mean radius of the DNA coil does not imply decompaction of the DNA coil, because the DNA still occupies only a limited portion of the confinement sphere. Results obtained with $|\sigma| > 0$ and $\rho = 0.65$, where the toroidal geometry is predominant, were therefore not displayed in Figs. 3 and 4, because $<R>$ is no longer a measure of the compaction of the DNA coil. Last but not least, it may be worth mentioning that toroidal DNA coils are also obtained in the absence of crowders and confinement when explicit and relatively strong attraction between DNA segments is plugged explicitly in a model of torsionally relaxed DNA (see for example Ref. (4)). This is of course completely different from what is observed in the fourth row of Fig. 5, where the toroidal geometry of the DNA coil is a direct consequence of the spherical geometry of the confinement chamber, which is specific to cocci. For non-spherical confinement chambers, like the capped cylinders specific to bacilli (including *E. coli*), DNA coils will rather relax towards more complex geometries which minimize both the bending energy of the DNA chain and the volume occupied by the coil.

**CONCLUSION**

The demixing of DNA from other macromolecules of the cytosol and the formation of plectonemes are two independent mechanisms, which are both able to compact the DNA coil.

In the present work, we investigated the interplay of these two mechanisms through coarse-grained modeling and Brownian dynamics simulations, with the goal of understanding how a topologically constrained DNA molecule compacts under the influence of non-interacting globular crowders. The model suggests that there exist three different regimes, depending on the superhelical density $\sigma$ of the DNA molecule and the effective volume ratio $\rho$ of the crowders:

(i) below a certain threshold for $|\sigma|$, the effects of the two mechanisms are additive, and the total compaction ratio of the DNA coil is the sum of the ratios due to demixing and to supercoiling. Compaction of supercoiled DNA is facilitated by the decrease of the diameter of the plectonemes with increasing values of $\rho$.

(ii) above this threshold for $|\sigma|$, the mean radius of the DNA coil ceases to decrease with increasing values of $|\sigma|$ because of the increasing stiffness of the plectonemes. For sufficiently large values of $\rho$, the DNA coil even decompacts upon increase of $|\sigma|$.

(iii) for values of $\rho$ just below the jamming threshold, the coil formed by topologically constrained DNA may adopt a non-spherical geometry, which represents some trade-off between the minimization of the bending energy and the minimization of the volume of the coil. For example, toroidal DNA coils were observed with the spherical confinement chamber.

According to the model, the threshold where the two mechanisms cease to have additive effects on the compaction of the DNA coil lies around $|\sigma| = 0.027$. This turns out to be almost exactly the value of the effective supercoil density which is observed in living cells ($\sigma = -0.025$), where DNA-binding proteins reduce the number of supercoils to approximately one half of the value in protein-free samples (90,91). Owing to the approximations of the model, such an exact correspondence is likely to be fortuitous, but it still suggests that both the additive and the non-additive regime may be relevant *in vivo*, because most biological functions rely on alternative winding and unwinding of the circular DNA (92,93). Moreover, the translational diffusion coefficient of macromolecules is much smaller in bacterial cells than in water and in eukaryotic cells (94), which indicates that the bacterial cytosol is indeed close to jamming. The regime predicted by the model, where the mean radius of the nucleoid increases with $|\sigma|$ instead of decreasing, and the abrupt change of its geometry very close to the jamming threshold, may consequently also be relevant for living cells. Finally, while the explicit modeling of topological domains is beyond the scope of this work, the results presented above support the hypothesis that differences in DNA concentration between different macro-domains (51-57) may indeed reflect differences in the level of superhelical density of the corresponding topological domains (58,59), provided that these domains are in the additive regime. An interesting related question is that of topological insulators and, more precisely, of the nature of the constraints these proteins must exert on the DNA duplexes they bind to in order to prevent the diffusion of DNA supercoils (95-98). Work in this direction is in progress.

More generally, it may be worth emphasizing that there exist, in addition to DNA demixing and supercoiling, several other mechanisms that may contribute to the compaction of the bacterial DNA and the formation of the nucleoid (for a recent review, see for example Ref. (4)). Each of these mechanisms, which have not been taken into account in the model proposed here, may interact additively or destructively with DNA demixing and supercoiling. Of particular interest is the action of nucleoid proteins, which can bridge (like H-NS), bend (like IHF, HU, and Fis), or wrap (like Dps) the DNA molecule (99). Cells lacking both HU and Fis have a large decondensed nucleoid (100), while the overproduction of H-NS leads instead to very compact nucleoids and may be lethal (101). Moreover, most nucleoid proteins, like Fis (102,103), HU (104-106), H-NS (102,107), and IHF (108) are capable of inducing gradual and strong DNA compaction *in vitro*, although at concentrations much larger than *in vivo* ones (109-111). It is known that nucleoid proteins interact with the underwinding of the DNA molecule (112) and are responsible for the fact that the number of supercoils in living cells is approximately one half of the value in protein-free samples (90,91). It may therefore be interesting in future work to introduce DNA-binding proteins in the model, as was done for example in Refs. (82-84), in order to investigate the interplay of DNA demixing and supercoiling with the binding of nucleoid proteins. In contrast, compaction of the DNA molecule in eukaryotic cells is primarily due to its wrapping around histone proteins, with supercoiling and crowding by non-binding macromolecules playing *a priori* a minor role compared to prokaryotes. This problem is therefore rather different from the formation of the bacterial nucleoid and its study requires the development of quite different models (see for example Ref. (113)).

## SUPPORTING MATERIAL

Model and Simulations section. Figures S1 to S3.

## SUPPORTING CITATIONS

Reference (114) appears in the Supporting Material.

# FIGURE CAPTIONS

**Figure 1** : Representative snapshots extracted from trajectories with $N=0$ ($\rho=0$) and $\sigma=0$, -0.027, -0.052, and -0.078. The solid red line connects the centers of successive DNA beads. Only one fourth of the confinement sphere is shown.

**Figure 2** : Representative snapshots extracted from trajectories with $N=1875$ ($\rho=0.61$) and $\sigma=0$, -0.027, -0.052, and -0.078. The solid red line connects the centers of successive DNA beads. Crowding spheres are colored in cyan and represented at 1/5 of their actual radius, in order that the DNA chain be seen through the layers of crowders. The confinement sphere is not shown.

**Figure 3** : Plot of the mean radius of the DNA coil, $<R>$, as a function of the opposite of the superhelical density of the DNA chain, $-\sigma$, for values of the effective crowder volume ratio $\rho$ ranging from 0 to 0.65. The lines are guides for the eyes. See text for explanations regarding the points that are missing for $\rho=0.65$.

**Figure 4** : Plot, as a function of the opposite of the superhelical density of the DNA chain, $-\sigma$, of the mean distance $<d>$ between the opposite strands of the plectonemes (top plot) and the mean contribution $<Wr/\Delta Lk>$ of the writhe to the linking number difference (bottom plot), for values of the effective crowder volume ratio $\rho$ ranging from 0 to 0.61. The lines are guides for the eyes.

**Figure 5** : Representative snapshots viewed either parallel (left column) or perpendicular (right column) to the main axis of inertia of the DNA chain. Trajectories were run with $N=1875$ ($\rho=0.61$) or $N=2000$ ($\rho=0.65$) and different values of $\sigma$. The solid red line connects the centers of successive DNA beads. Crowding spheres are not shown. The blue circle shows the limits of the confinement sphere. Note the approximate spherical geometry of the DNA coil in the three top lines and its approximate toroidal geometry in the bottom line.

**FIGURE 1**

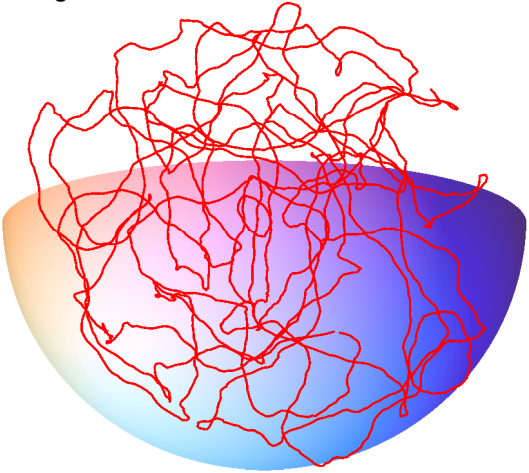
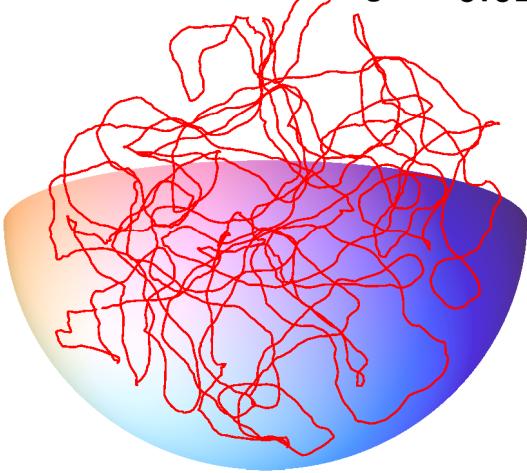
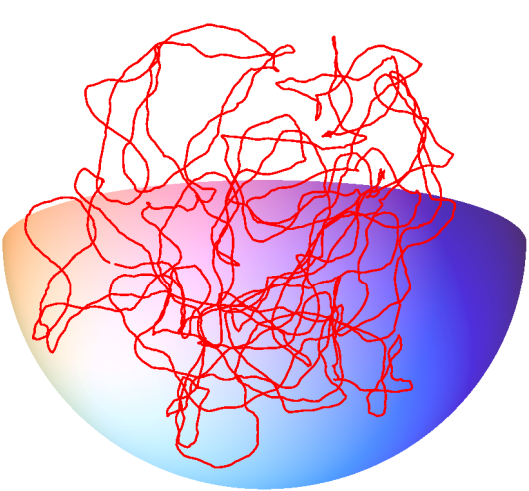
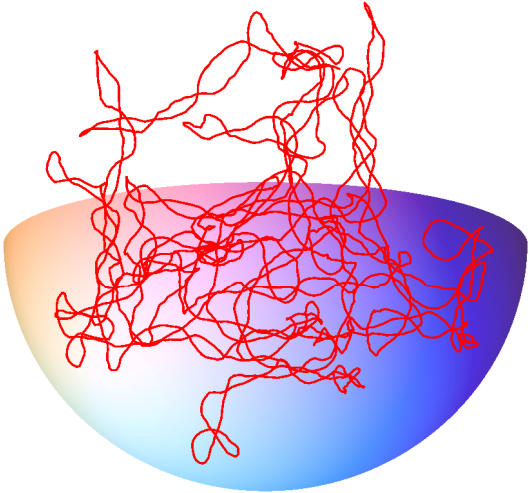

σ = 0

σ = −0.027

σ = −0.052

σ = −0.078

**FIGURE 2**

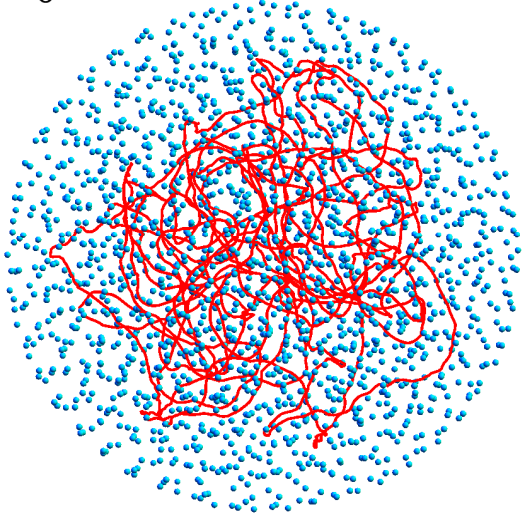
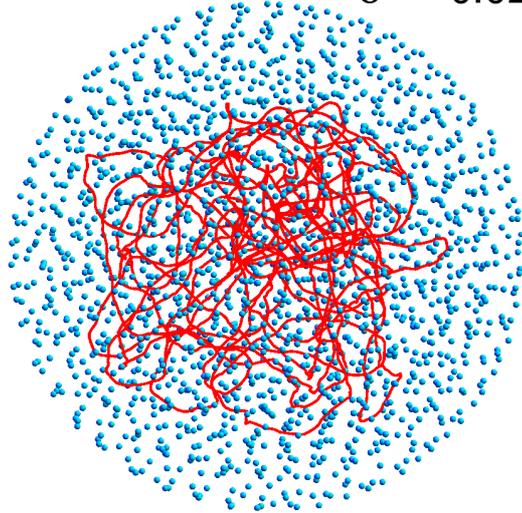
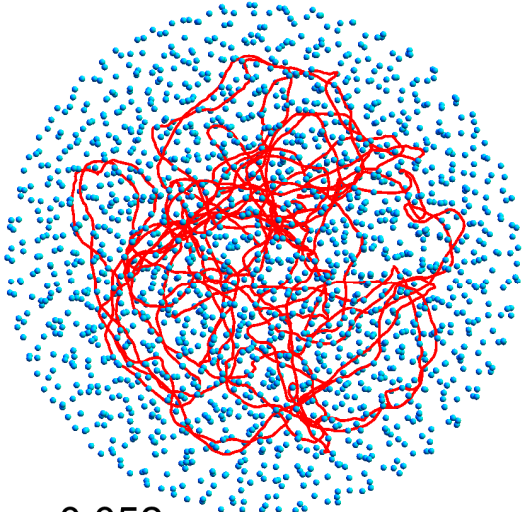
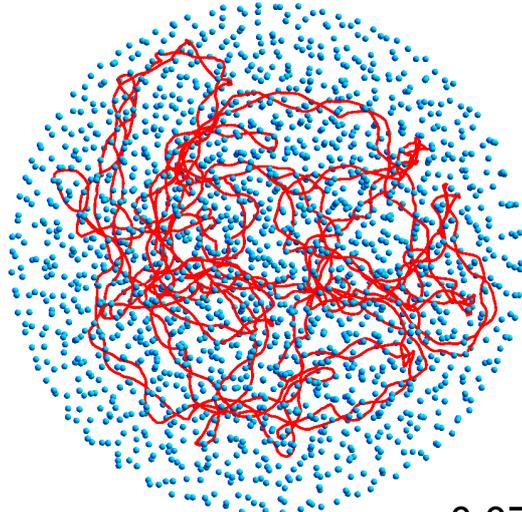

**FIGURE 3**

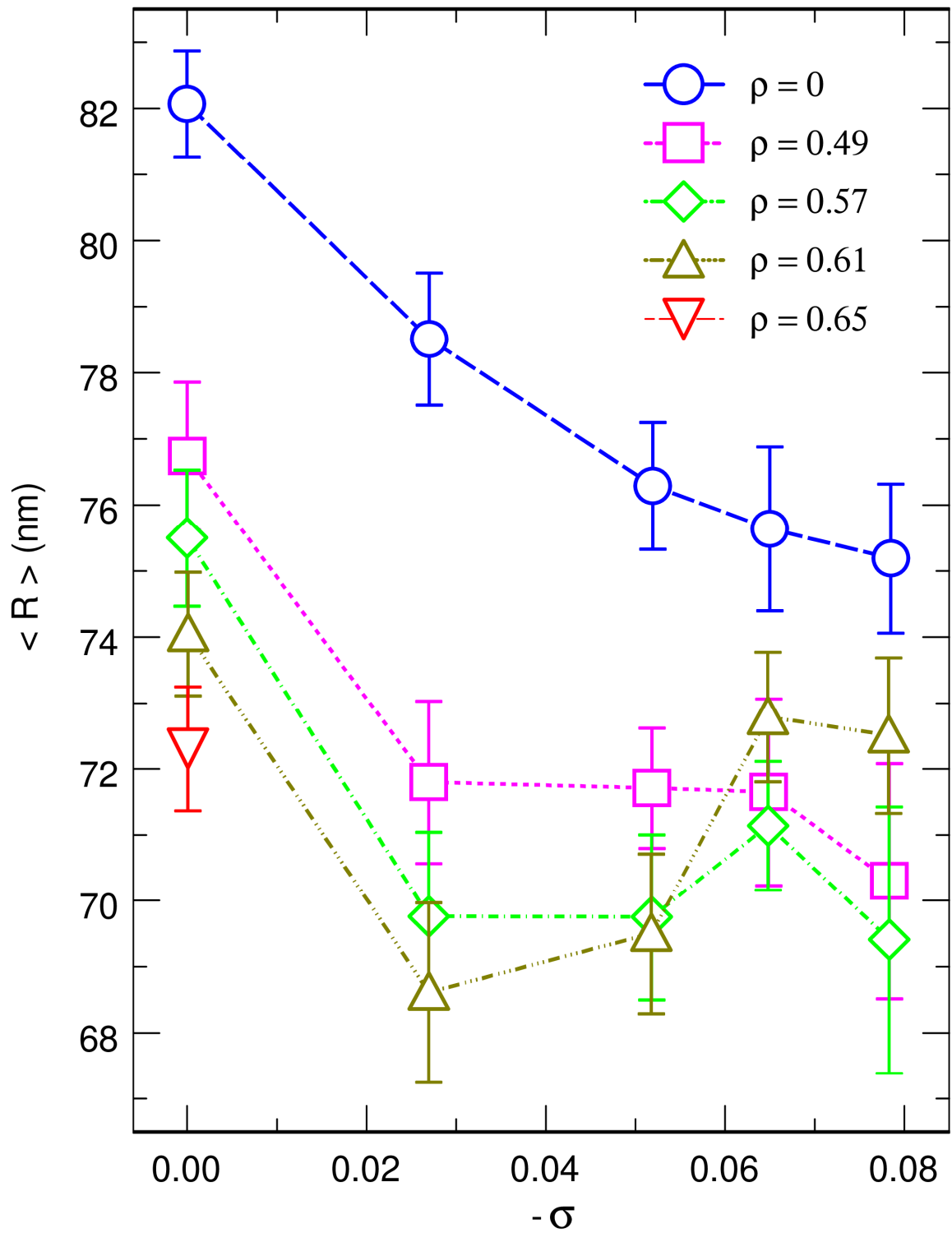

**FIGURE 4**

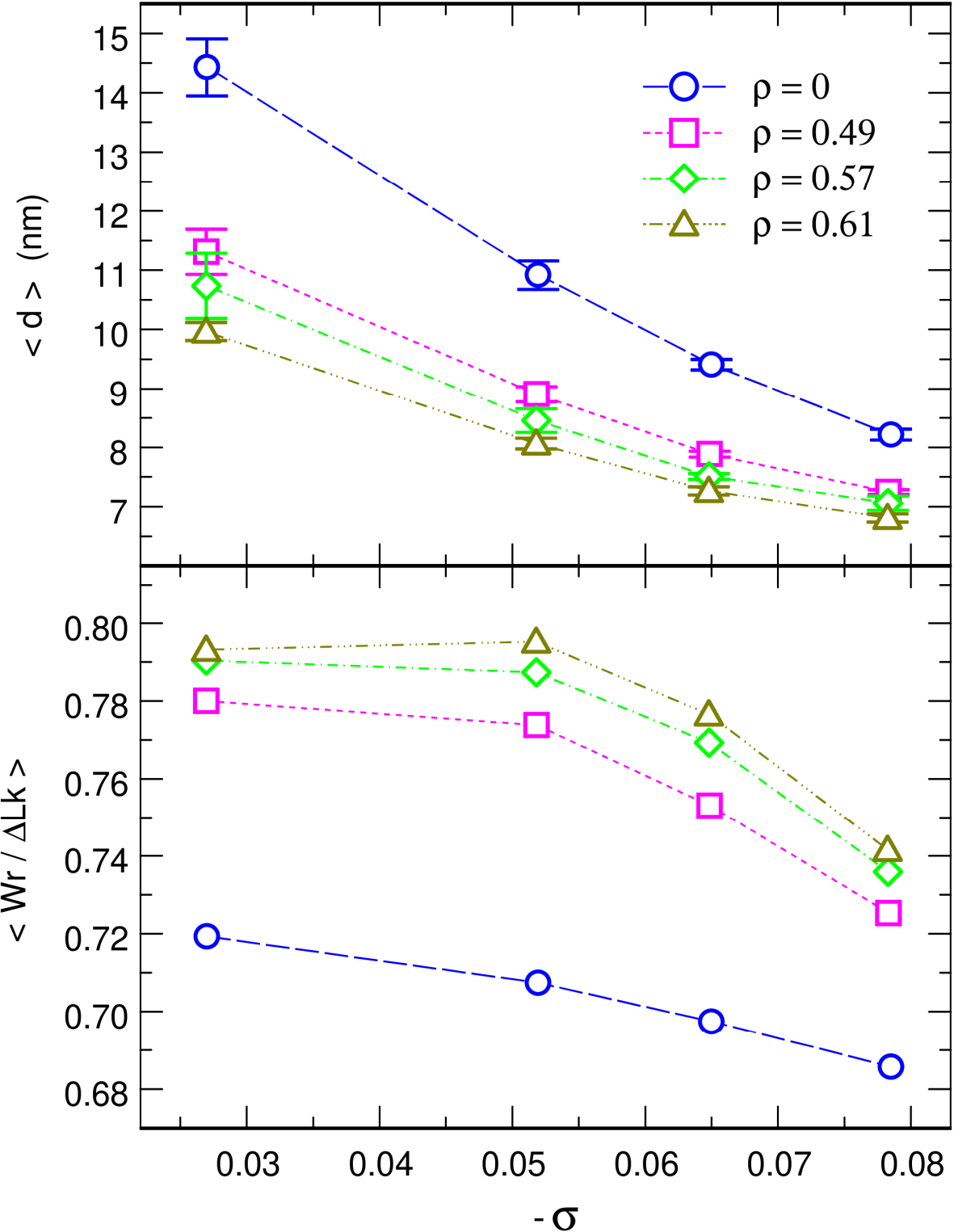

**FIGURE 5**

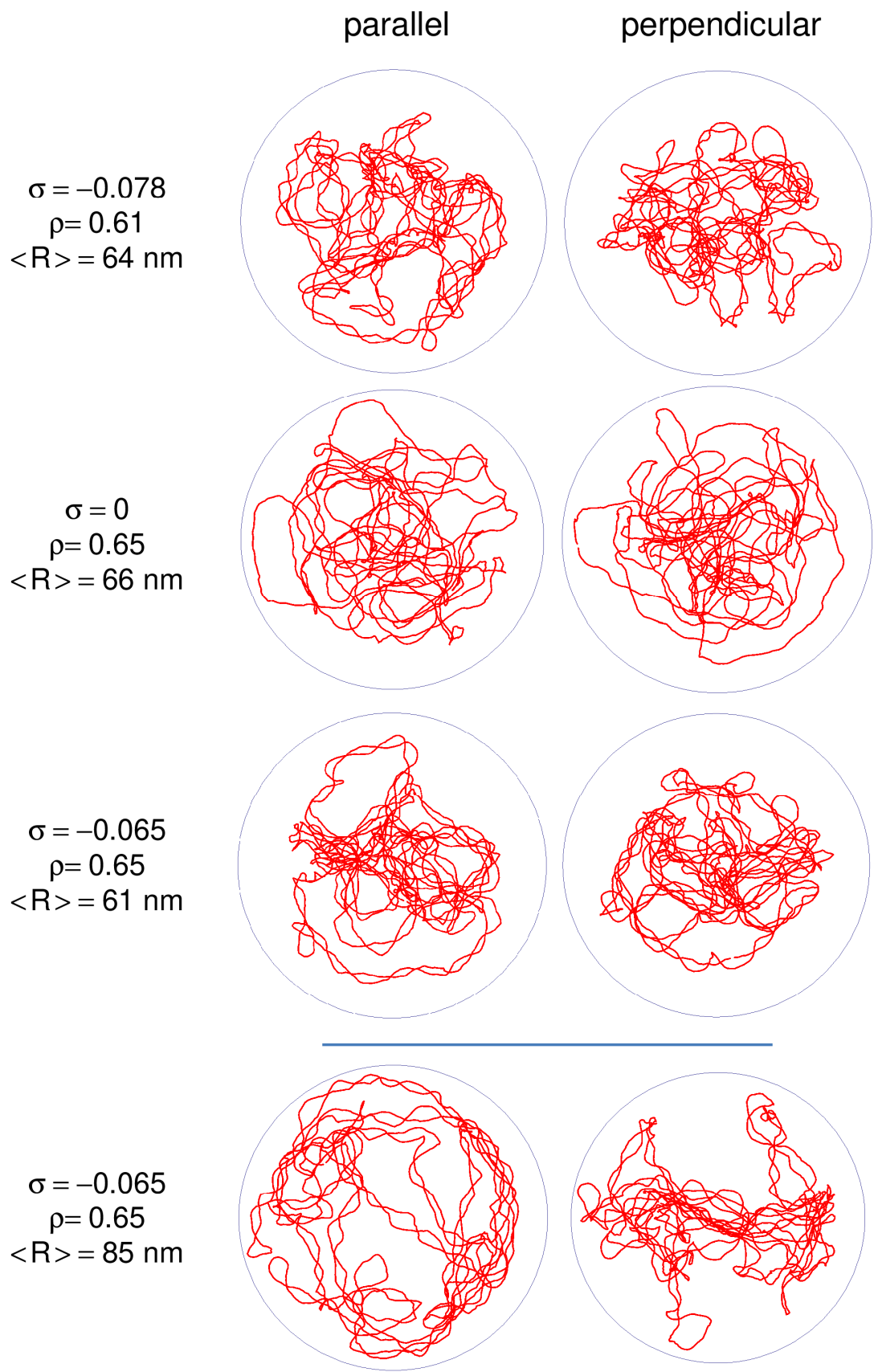

**Bacterial nucleoid: Interplay of DNA demixing and supercoiling**

- Supporting Material –

M. Joyeux

*Laboratoire Interdisciplinaire de Physique,*

*CNRS and Université Grenoble Alpes,*

*Grenoble, France*

**MODEL AND SIMULATIONS**

Temperature $T$ is assumed to be 298 K throughout the study. The model consists of a circular chain of $n = 2880$ beads with radius $a = 1.0$ nm separated at equilibrium by a distance $l_0 = 2.5$ nm (the genomic DNA) enclosed in a large confinement sphere of radius $R_0 = 120$ nm (the cell), together with $N$ spheres of radius $b = 7.4$ nm (the crowding globular macromolecules). Two beads represent 15 DNA base pairs. The contour length of the DNA chain and the volume of the confinement sphere correspond approximately to 1/200th of the values for *E. coli* cells, so that the DNA concentration of the model is close to the physiological value (about 10 mM). The potential energy of the system, $E_{\text{pot}}$, consists of four terms

$$E_{\text{pot}} = V_{\text{DNA}} + V_{\text{DNA/C}} + V_{\text{C/C}} + V_{\text{wall}} \; , \tag{S1}$$

which describe the internal energy of the DNA molecule, DNA-crowder interactions, crowder-crowder interactions, and the repulsive potential that maintains DNA beads and crowding spheres inside the confinement sphere, respectively. The internal energy of the DNA chain is further written as the sum of 4 contributions

$$V_{\text{DNA}} = \frac{h}{2}\sum_{k=1}^{n}(l_k - l_0)^2 + \frac{g}{2}\sum_{k=1}^{n}\theta_k^2 + \frac{\tau}{2}\sum_{k=1}^{n}(\Phi_{k+1} - \Phi_k)^2 + q^2\sum_{k=1}^{n-2}\sum_{K=k+2}^{n}H(\|\mathbf{r}_k - \mathbf{r}_K\| - 2a) \; , \tag{S2}$$

where

$$H(r) = \frac{1}{4\pi\varepsilon r}\exp\left(-\frac{r}{r_D}\right) \; , \tag{S3}$$

which describe the stretching, bending, torsional, and electrostatic energy of the DNA chain, respectively. $\mathbf{r}_k$ denotes the position of DNA bead $k$, $l_k = \|\mathbf{r}_k - \mathbf{r}_{k+1}\|$ the distance between two successive beads, and $\theta_k = \arccos((\mathbf{r}_k - \mathbf{r}_{k+1})(\mathbf{r}_{k+1} - \mathbf{r}_{k+2})/(\|\mathbf{r}_k - \mathbf{r}_{k+1}\|\|\mathbf{r}_{k+1} - \mathbf{r}_{k+2}\|))$ the angle formed by three successive beads. The stretching energy is a computational device without biological meaning, which is aimed at avoiding a rigid rod description. The stretching force constant $h$ was set to $h = 100\, k_B T / l_0^2$ to insure that the variations of the distance between successive beads remain small enough (1). In contrast, the bending rigidity was obtained from the known persistence length of DNA, $\xi = 50$ nm, according to $g = \xi\, k_B T / l_0 = 20\, k_B T$. The torsion contribution is borrowed from Ref. (2) and torsional forces and momenta are computed as described therein. $\Phi_{k+1} - \Phi_k$ denotes the rotation of the body-fixed frame $(\mathbf{u}_k, \mathbf{f}_k, \mathbf{v}_k)$ between DNA beads $k$ and $k+1$. The value of the torsional rigidity, $\tau = 25\, k_B T$, was obtained by imposing that the writhe contribution $Wr$ accounts for approximately 70% of the linking number difference $\Delta Lk$ at equilibrium (3), see Fig. 4. The value of the torsional rigidity used in the simulations is close to the value of the bending rigidity, which agrees with experimental findings (3). Finally, the electrostatic energy of the DNA chain is written as a sum of repulsive Debye-Hückel terms with hard core. $\varepsilon = 80\, \varepsilon_0$ denotes the dielectric constant of the buffer and $r_D = 1.07$ nm the Debye length inside the buffer. This value of the Debye length corresponds to a concentration of monovalent salt of 100 mM, which is the value that is generally assumed for the cytoplasm of bacterial cells. $q$ is the value of the electric charge, which is placed at the centre of each DNA bead

$$q = -\frac{l_0 \bar{e}}{\ell_B} \approx -3.5\, \bar{e}\,, \tag{S4}$$

where $\bar{e}$ is the absolute charge of the electron and $\ell_B = 0.7$ nm the Bjerrum length of water. In Eq. (S4), $\bar{e}/\ell_B$ is the net linear charge density along a DNA molecule immersed in a buffer with monovalent cations derived from Manning's counterion condensation theory (4,5). Note that electrostatic interactions between nearest neighbours are not included in Eq. (S2) because it is considered that they are already accounted for in the stretching and bending terms.

DNA-crowder and crowder-crowder interactions are similarly expressed as sums of Debye-Hückel potentials with hard cores

$$V_{\text{DNA/C}} = q^2 \sum_{k=1}^{n} \sum_{j=1}^{N} H(\|\mathbf{r}_k - \mathbf{R}_j\| - a - b)$$

$$V_{\text{C/C}} = q^2 \sum_{j=1}^{N-1} \sum_{J=j+1}^{N} H(\|\mathbf{R}_j - \mathbf{R}_J\| - 2b) ,$$

(S5)

where $\mathbf{R}_j$ denotes the position of crowding sphere $j$, and a charge $q$ is placed at the centre of each crowding sphere, as for DNA beads. The repulsion potential between a DNA bead and a crowding sphere is therefore the median of the repulsion potential between two DNA beads and the repulsion potential between two crowding spheres. According to previous work (6,7), strong compaction of torsionally relaxed DNA chains is consequently expected for effective volume fractions of the crowders

$$\rho = \frac{N(b + \Delta b)^3}{R_0^3} ,$$

(S6)

in the range $0.60 \leq \rho \leq 0.70$, that is close to the jamming threshold for hard spheres. In Eq. (S6), $b + \Delta b$ represents the effective radius of the crowding spheres, that is half the distance between the centres of two spheres at which the electrostatic repulsion energy is equal to the thermal energy $k_B T$. Numerical values of the parameters reported above lead to $\Delta b = 0.865$ nm, so that $\rho \approx 0.49$, 0.57, 0.61, and 0.65 for $N = 1500$, 1750, 1875, and 2000, respectively.

Finally, $V_{\text{wall}}$ is written in the form

$$V_{\text{wall}} = \zeta \left( \sum_{k=1}^{n} f(\|\mathbf{r}_k\|) + \sum_{j=1}^{N} f(\|\mathbf{R}_j\|) \right) ,$$

(S7)

where the repulsive force constant $\zeta$ is set to $1000 k_B T$ and the function $f(r)$ is defined according to

if $r \leq R_0$ : $f(r) = 0$

if $r > R_0$ : $f(r) = \left( \frac{r}{R_0} \right)^6 - 1$ .

(S8)

The dynamics of the system was investigated by integrating numerically overdamped Langevin equations. Practically, the updated positions and torsion angles at time step $i+1$ are computed from the positions and torsion angles at time step $i$ according to

$$\mathbf{r}_k^{(i+1)} = \mathbf{r}_k^{(i)} + \frac{\Delta t}{6\pi\eta a}\mathbf{f}_k^{(i)} + \sqrt{\frac{2 k_B T \Delta t}{6\pi\eta a}}\, x_k^{(i)}$$

$$\mathbf{R}_j^{(i+1)} = \mathbf{R}_j^{(i)} + \frac{\Delta t}{6\pi\eta b}\mathbf{F}_j^{(i)} + \sqrt{\frac{2 k_B T \Delta t}{6\pi\eta b}}\, X_j^{(i)} \qquad (S9)$$

$$\Phi_k^{(i+1)} = \Phi_k^{(i)} + \frac{\tau \Delta t}{4\pi\eta a^2 l_0}(\Phi_{k+1}^{(i)} - 2\Phi_k^{(i)} + \Phi_{k-1}^{(i)}),$$

where $\mathbf{f}_k^{(i)}$ and $\mathbf{F}_j^{(i)}$ are vectors of inter-particle forces arising from the potential energy $E_{pot}$, $T = 298$ K is the temperature of the system, $x_k^{(i)}$ and $X_j^{(i)}$ are vectors of random numbers extracted from a Gaussian distribution of mean 0 and variance 1, $\eta = 0.00089$ Pa s is the viscosity of the buffer at 298 K, and $\Delta t = 10$ ps is the integration time step. After each integration step, the position of the centre of the sphere was slightly adjusted so as to coincide with the centre of mass of the DNA molecule. Trajectories were integrated for 10 ms for $N = 0$, 20 ms for $N = 1500$ and $N = 1750$, 30 ms for $N = 1875$, and up to 80 ms for $N = 2000$.

The twist difference $\Delta T\mathrm{w}$, the writhe $Wr$, and the linking number difference $\Delta Lk$, of the DNA chain were computed at regular time intervals according to (2)

$$\Delta T\mathrm{w} = \frac{1}{2\pi}\sum_{k=1}^{n}(\Phi_{k+1} - \Phi_k)$$

$$Wr = \frac{1}{4\pi}\sum_{k=1}^{n}\sum_{K\neq k}\frac{[(\mathbf{r}_{k+1} - \mathbf{r}_k)\times(\mathbf{r}_{K+1} - \mathbf{r}_K)]\cdot(\mathbf{r}_k - \mathbf{r}_K)}{|\mathbf{r}_k - \mathbf{r}_K|^3} \qquad (S10)$$

$$\Delta Lk = \Delta T\mathrm{w} + Wr,$$

The superhelical density $\sigma$ was subsequently estimated from

$$\sigma = \frac{\Delta Lk}{Lk_0}, \qquad (S11)$$

where the linking number $Lk_0 = 7.5 n/10.5$ is the ratio of the formal number of base pairs of the DNA chain and the mean number of base pairs per turn of the torsionally relaxed double helix. $Lk_0 \approx 2057$ for $n = 2880$ and $Lk_0 \approx 143$ for $n = 200$.

# SUPPORTING REFERENCES

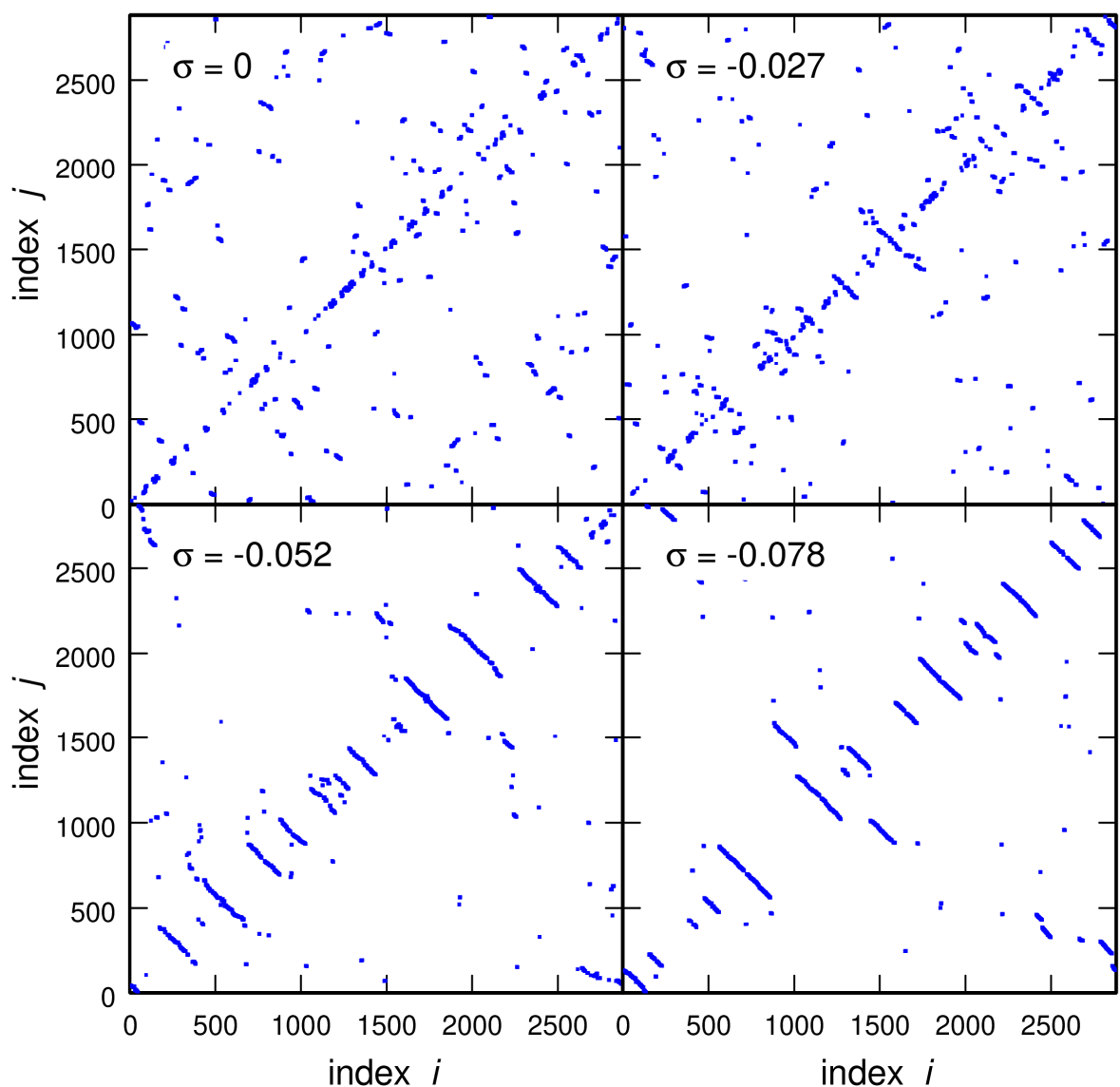

**Figure S1 :** Plot of the index *j* of the bead located closest to bead *i* for the equilibrated DNA conformations shown in Fig. 1. The segments of points parallel to the anti-diagonal denote the presence of plectonemes.

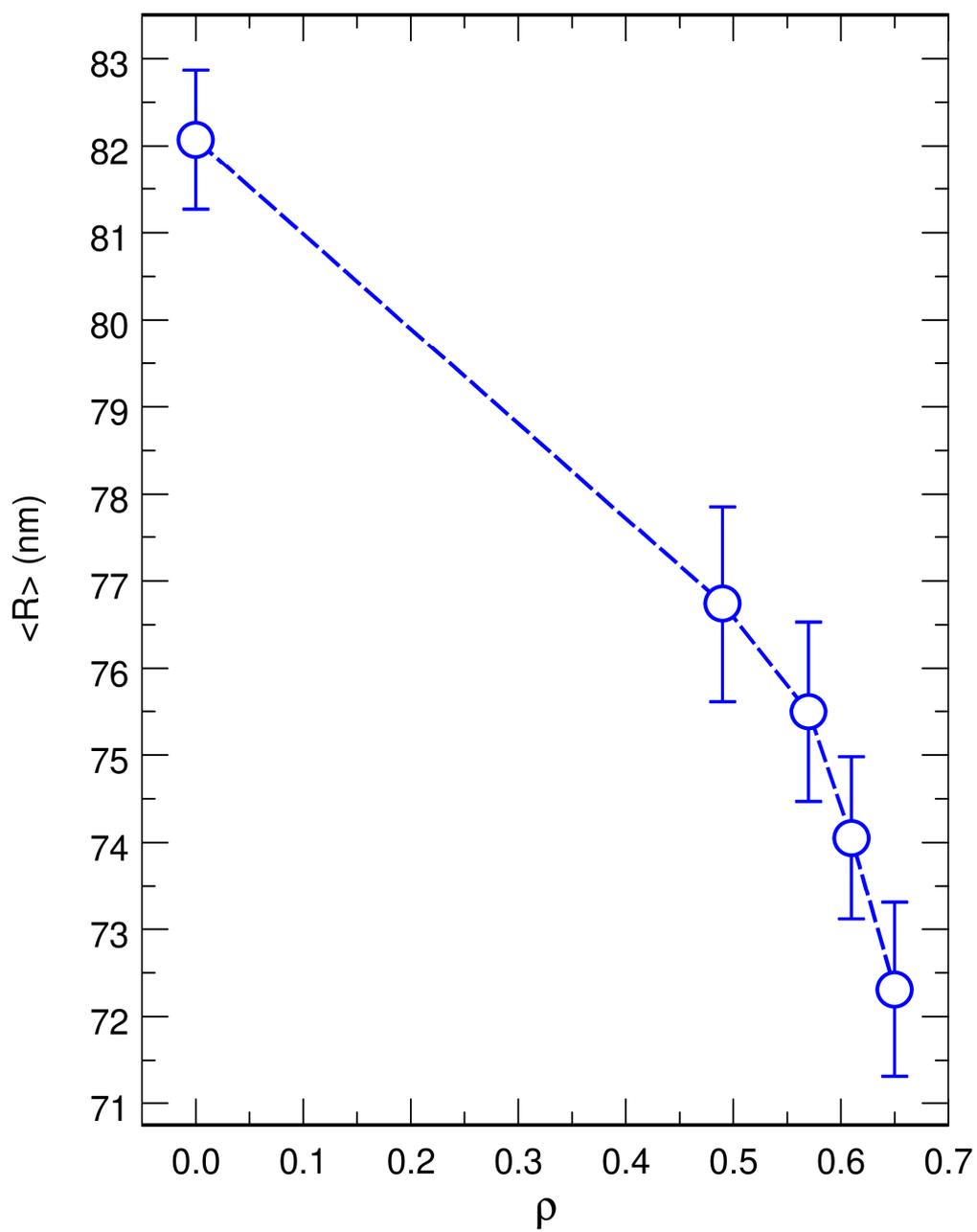

**Figure S2** : Plot, as a function of the effective crowder volume ratio $\rho$, of the mean radius of the DNA coil $<R>$ for torsionally relaxed DNA chains ($\sigma = 0$).

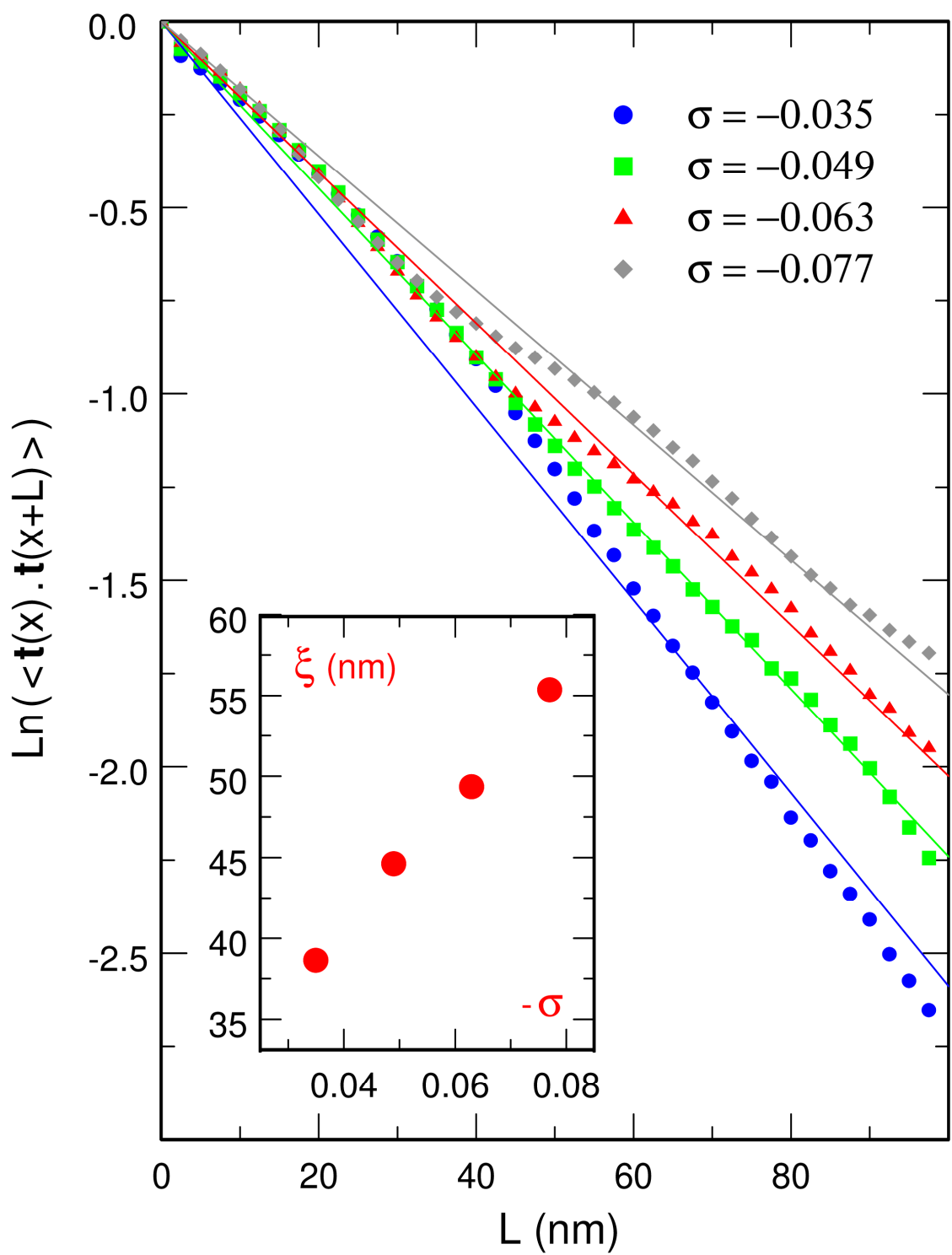

**Figure S3** : **(Main plot)** Plot of the logarithm of the directional correlation function of the segments of the mid curve of plectonemes for different values of the superhelical density $\sigma$ of a DNA chain with 200 beads (1500 bp). The solid lines show the results of linear fits according to $\mathrm{Ln}(<\mathbf{t}(x)\cdot\mathbf{t}(x+L)>) = -L/\xi$, where $\mathbf{t}(x)$ is the tangent to the chain at position $x$, $L$ the distance between the two segments, and $\xi$ the fitted persistence length **(Insert)** Evolution of $\xi$ as a function of $-\sigma$.